\RequirePackage{fix-cm}
\documentclass[smallextended]{svjour3}       
\RequirePackage{amsmath,amsfonts,amssymb}
\RequirePackage[numbers]{natbib}
\RequirePackage{graphicx}
\usepackage{enumerate}
\usepackage{lineno}
\usepackage{array}
\usepackage{subfigure}
\usepackage{color}

\usepackage{graphicx}

\begin{document}

\title{Most Powerful Test Sequences with Early Stopping Options}

\titlerunning{mpact of Early Stopping Options}        

\author{Sergey Tarima         \and
        Nancy Flournoy
}

\authorrunning{Flournoy and Flournoy} 

\institute{Institute for Health and Society, Medical College of Wisconsin, 8701 Watertown Plank Rd, Wauwatosa, WI, 53226
           \and
Department of Statistics, University of Missouri, 600 S State St., Apt. 408
Bellingham, WA 98225}

\date{Received: date / Accepted: date}

\maketitle

\begin{abstract}
 Sequential likelihood ratio testing is found to be most powerful in  sequential studies with early stopping rules when grouped data come from the one-parameter exponential family.    
First, to obtain this elusive result, the probability measure of a group sequential design is constructed with support for all possible outcome events, as is useful for designing an experiment prior to having data.  This construction identifies  impossible events that  are not part of the support.  The overall  probability distribution is dissected  into stage specific  components. These components are sub-densities of interim test statistics first described by Armitage, McPherson and Rowe (1969) that are commonly used to create stopping boundaries given an $\alpha$-spending function and a set of interim analysis times.  Likelihood expressions conditional on reaching a stage are given to connect pieces of the  probability anatomy together.

The reduction of the support  caused by the adoption of an early stopping rule induces sequential truncation (not nesting) in the probability distributions of  possible events. 
   Multiple testing induces mixtures on the adapted support. Even asymptotic distributions of inferential statistics are mixtures of truncated distributions. In contrast to the classical result on local asymptotic normality (Le Cam 1960), statistics that are asymptotically normal without stopping options have asymptotic distributions that are  mixtures of truncated normal distributions under local alternatives with stopping options; under fixed alternatives, asymptotic distributions of test statistics are degenerate.

\keywords{Adaptive designs \and adapted support \and group sequential designs \and local asymptotics \and interim hypothesis testing \and likelihood ratio tests}
\end{abstract}

\section{Introduction} \label{Introduction}
We define a \emph{sequential experiment} to be one in which the decision to stop collecting data is based on data collected previously in the study.
\citet{wetherill1986} emphasize that "\emph{two aspects of a sequential procedure must be clearly distinguished, the stopping rule, and the manner in which inferences are made once observations are stopped. $\ldots$ in the design problem, it is important to know how probable are various possible results."} We distinguish the probability framework underpinning these two activities, and also a third - the probability framework underlying interim hypothesis tests.



\citet{dodge1929} proposed the first known sequential test procedure in which a decision to stop or continue collecting data was based on prior data, recognizing that decisions to stop or continue a trial  made  based on prior observations could substantially reduce the expected numbers of required subjects; it was a two-stage design.  \citet{bartky1943} devised a multiple sequential testing procedure for binomial data based on \citet{neyman1933}'s likelihood ratio test  that \citet{wald1947} cites as a "forerunner" to his more general \emph{sequential probability ratio test} (SPRT)  procedure in which the probabilities of type I and II errors are controlled.  Extensions with stopping decisions based on groups of subjects  [\emph{Group Sequential Designs (GSDs)}] are given in  \citet{jennison1999}.

\citet{neyman1933} show that likelihood ratio tests are most powerful for testing a simple null versus a simple alternative hypotheses. In \citet{ferguson2014mathematical}, the Karlin-Rubin theorem is viewed as an extension of Neyman-Pearson approach to most powerful testing of composite hypotheses. The Karlin-Rubin theorem applies to the one-dimensional exponential family. With sequential stopping options, some elements of the sample space become impossible, which changes the distributions of statistics. Sufficient statistics become dependent on the random sample size; see \citet{blackwell1947}. If statistics belong to   the exponential family without early stopping options,  then  they belong to a curved exponential family when exposed to early stopping options (\citet{efron1975, liu1999,liu2006}). Section \ref{mostpower} shows that despite being from a  curved exponential family, the sequential tests based on likelihood ratios continue to be most powerful for any $\alpha$-spending function.  

\subsection{Notation and A Simple Example}\label{sec:ex_simple}
This example demonstrates a couple important repercussions of sequential stopping rules: \begin{itemize}
\setlength\itemsep{.0em}
 \item The support is reduced, 
 \item  Bivariate normal random variables become non-observable;  the observable bivariate random variable is a mixture of truncated normal random variables, 
 \end{itemize} 
Let $X_1$ and $X_2$ be  $N(\theta,1)$ random variables with an unknown location parameter $\theta$.

\subsubsection{Non-sequential experiments}
If $X_1=x_1$ alone is observed, the log-likelihood  $$l(\theta|X_1=x_1)=-2\log \sqrt{2\pi} -0.5 (x_1-\theta)^2$$ is maximized at $\widehat\theta_1=x_1$.

If both $X_1=x_1$ and $X_2=x_2$ are 
 observed independently, 
 the random variable $(X_1,X_2)$ is defined on the probability space $\left(R^2,{\cal{B}}, P\right)$, where $R^2$ is the sample space [$R=\left(-\infty,\infty\right)$], 
${\cal{B}}$ is the Borel $\sigma$-algebra on $R^2$ and $P$ is  a bivariate normal distribution with mean vector $(\theta,\theta)$, units variances   and zero correlation.

 The log-likelihood function is
 $$l(\theta|X_1=x_1, X_2=x_2) = -2 \log \sqrt{2\pi} -0.5 (x_1-\theta)^2 -0.5 (x_2-\theta)^2$$ and the maximum likelihood estimator (MLE) of $\theta$ is $\widehat\theta_2=(x_1+x_2)/2$. 

\subsubsection{Sequential experiments: likelihood, support and probability measures }\label{sec:seq_exp}
What happens in \textit{sequential settings} when $X_2$ is only observed if $X_1<2.18$? Let $D$ denote
the random stopping stage. Then $D = 1+I(X_1< 2.18)$, where $I(\cdot)$ is an indicator function. In this simple example, $D$ is also the random sample size.

The joint distribution of (\textbf{X},D) and the marginal distribution of \textbf{X} are the same:
\begin{align}\label{eq:jtden}
f_{\textbf{X}}\left(\textbf{x}|\theta \right)=f_{\textbf{X},D}\left(\textbf{x},d|\theta \right) &= \phi(x_1-\theta) \left[\phi(x_2-\theta)\right]^{I\left(d=2\right)}\\&= [\phi(x_1-\theta)]^{I(d=1)} \left[\phi(x_1-\theta)\phi(x_2-\theta)\right]^{I\left(d=2\right)},\notag 
\end{align}
where $\phi(\cdot)$ is the standard normal density. The representation of $f_{\textbf{X}}\left(\textbf{x}|\theta \right)$ in the first line of \eqref{eq:jtden} partitions the density according to data collection stages, while the representation in the second line partitions the density according to stopping stages. In canonical form,
\begin{eqnarray}f_{\textbf{X},D}\left(\textbf{x},d|\theta \right) 
&=&
h(\mathbf{x})\exp \Big(\left[x_1 + I(d=2)x_2\right]\theta  - \left[1+I(d=2)\right]\frac{\theta^2}{2}\Big), \label{marginal_density}
\end{eqnarray}
where $h(\mathbf{x}) = \exp \Big( -\frac{1}{2} \left[x_1^2 + I(d=2)x_2^2 + I(d=2) \log \left(\sqrt{\pi}\right) +\log \left(\sqrt{\pi} \right)\right]
\Big)$. Thus, the  density (\ref{marginal_density}) belongs to the curved exponential family with a sufficient statistic $$\left(\sum_{k=1}^d x_i, d\right) = \left[x_1 + I(d=2)x_2,\, 1+I(d=2)\right].$$ Curved exponential families were defined by \citet{efron1975} and the sufficient statistic with a random number of summands $\sum_{k=1}^Dx_k$ was derived by \citet{blackwell1947}.  Probability distribution (\ref{marginal_density}) is a special case of the exponential family derived in \citet{liu2006} [see their formula (2.6)].  A more general probability  distribution of $\mathbf{X}$ is presented in Section \ref{sec:densities}.

The log-likelihood function 
\begin{align*}
 l(\theta|X_1=x_1, X_2=x_2, D=d) &= \begin{cases}
[\phi(x_1-\theta)]^{I(d=1)}  &\textrm{ if } D=1, \\ \left[\phi(x_1-\theta)\phi(x_2-\theta)\right]^{I\left(d=2\right)} & \textrm{ if }  D=2
\end{cases}
\end{align*}
 is maximized at 
$\widehat\theta = I(d=1)x_1 + I(d=2)(x_1+x_2)/2.$
Consequently, the score function, the MLE and the observed information are the same as for the non-sequential experiment. What changes?

The joint support of $X_1$ and $X_2$ changes because $X_2$ becomes  impossible (not just missing) when $X_1\ge 2.18$. The random variable $(X_1,X_2)$ is non-observable when $X_1\ge 2.18$ and the joint distribution of $X_1$ and $X_2$ is therefore truncated and not normal. Formally,
the support for joint density can be decomposed into support for the experiment stopping with $X_1$ and support for the experiment continuing to observe $X_2$:  $${\cal{T}}=\{x_1 \ge 2.18 \}\ \cup \ \{ \{x_1 < 2.18\} \cap \{x_2  \in R\} \}\ \subset R^2.$$ 
This ${\cal{T}}$ is a special case of the support formalized in \citet{liu2006}.
If $A \in {\cal{T}}$,  then
\begin{eqnarray}
P_{(D)} = \text{Pr}((X_1,X_2) \in A) &=& \text{Pr}(X_1 \ge 2.18)\text{Pr}(X_1 \in A|X_1 \ge 2.18) \notag \\ 
&+& 
\text{Pr}(X_1 < 2.18)\text{Pr}((X_1, X_2) \in A|X_1 < 2.18)
\end{eqnarray}
is a probability measure on the $\sigma$-algebra $\sigma({\cal{T}})$. Thus, the observable random variable
\begin{align*}
 \mathbf{X}_{{\cal{T}}} = \begin{cases}
X_1 &\textrm{ if } $D=1$ \\ (X_1,X_2) & \textrm{ if }  $D=2$.
\end{cases}
\end{align*}
 is defined on the probability space $\left({\cal{T}}, \sigma({\cal{T}}), P_{(D)}\right).$

In contrast to $(X_1,X_2)$, $\mathbf{X}_{{\cal{T}}}$ is  observable for this sequential experiment.
The  MLE,  is a random variable defined on $\left({\cal{T}}, \sigma({\cal{T}}), P_{(D)}\right)$ and its probability distribution is a mixture of the left truncated normal random variable $\{X_1|D=1\}$ and an average of the right truncated normal random variable $\{X_1|D=2\}$ and the normal random variable $X_2$. 

\begin{figure}[bt!]
\centering
\subfigure[$x_1:x_1\ge 2.18$, $\theta=0$]{\includegraphics[width = 1.4in]{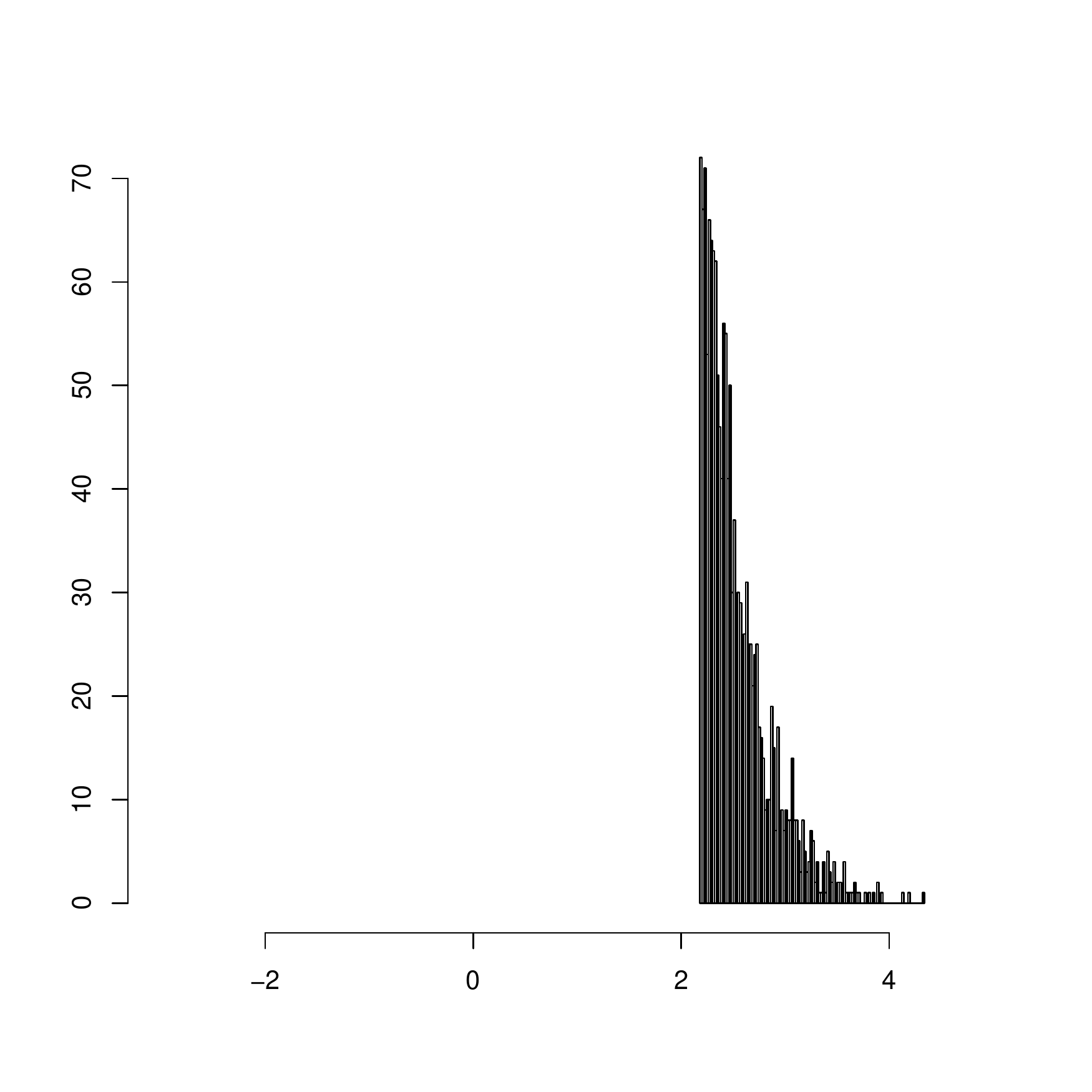}} \quad
\subfigure[$\frac{x_1+x_2}{2}:x_1< 2.18$, $\theta=0$]{\includegraphics[width = 1.4in]{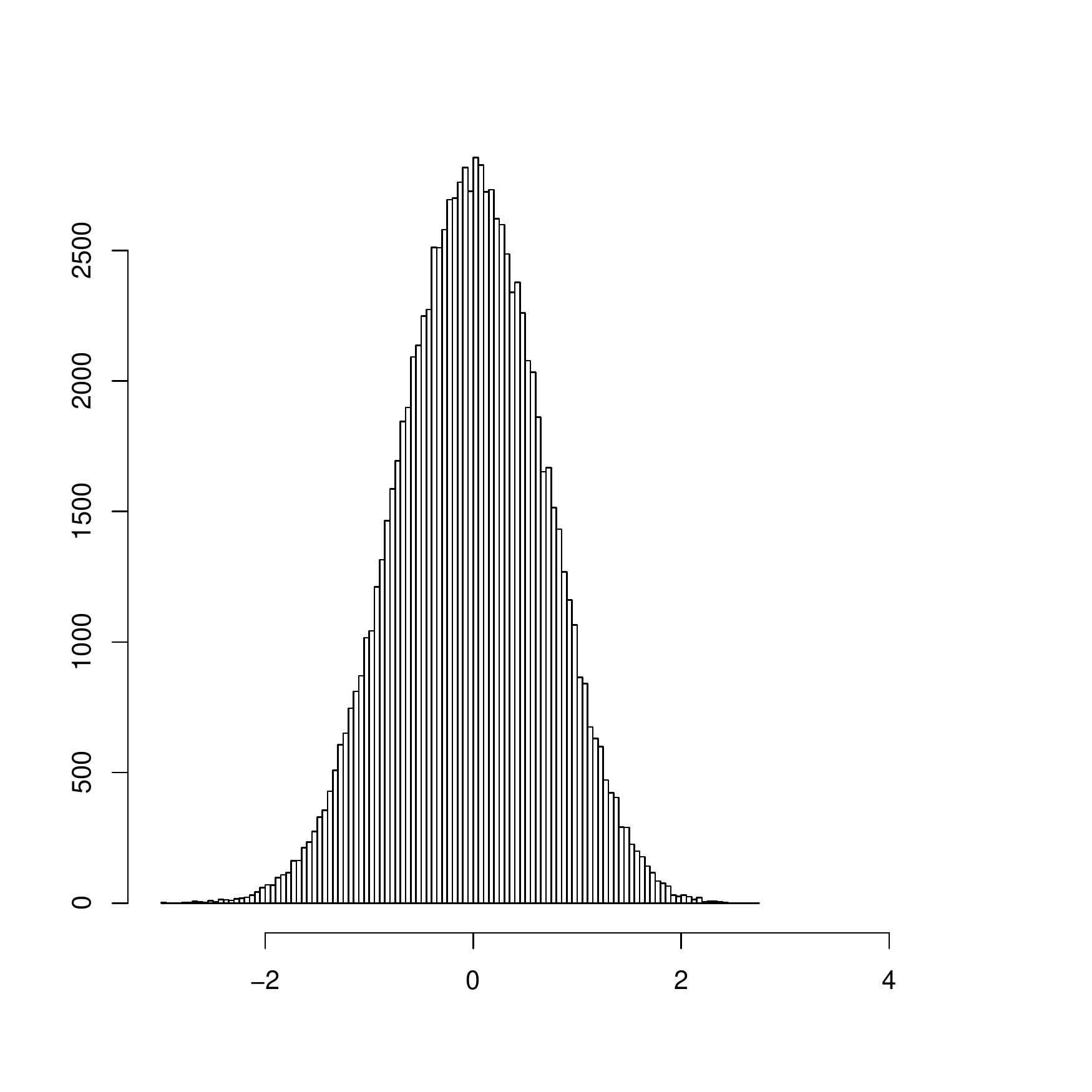}} \quad
\subfigure[$\widehat\theta$, $\theta=0$]{\includegraphics[width = 1.4in]{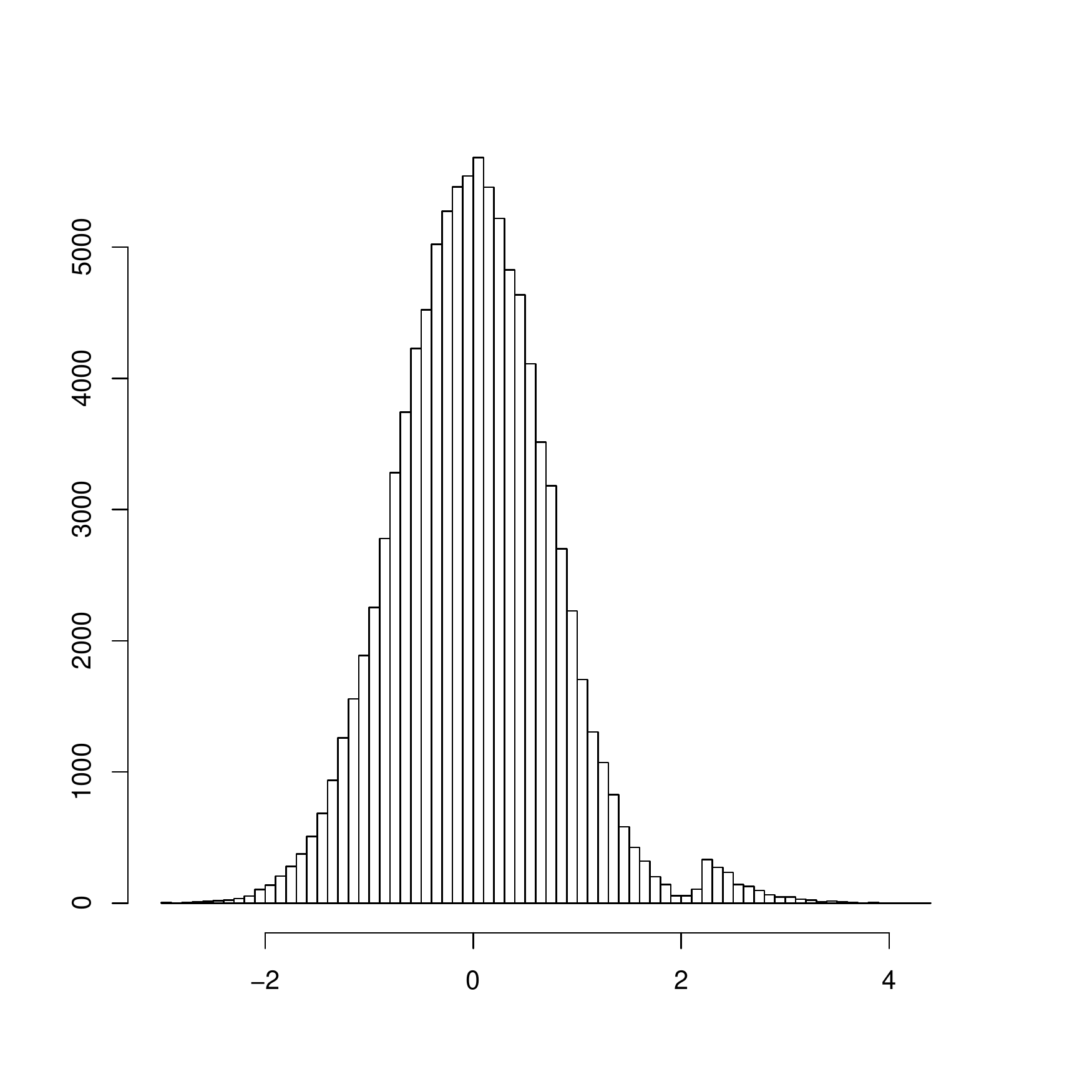}} \\
\subfigure[$x_1:x_1\ge 2.18$, $\theta=2.18$]{\includegraphics[width = 1.4in]{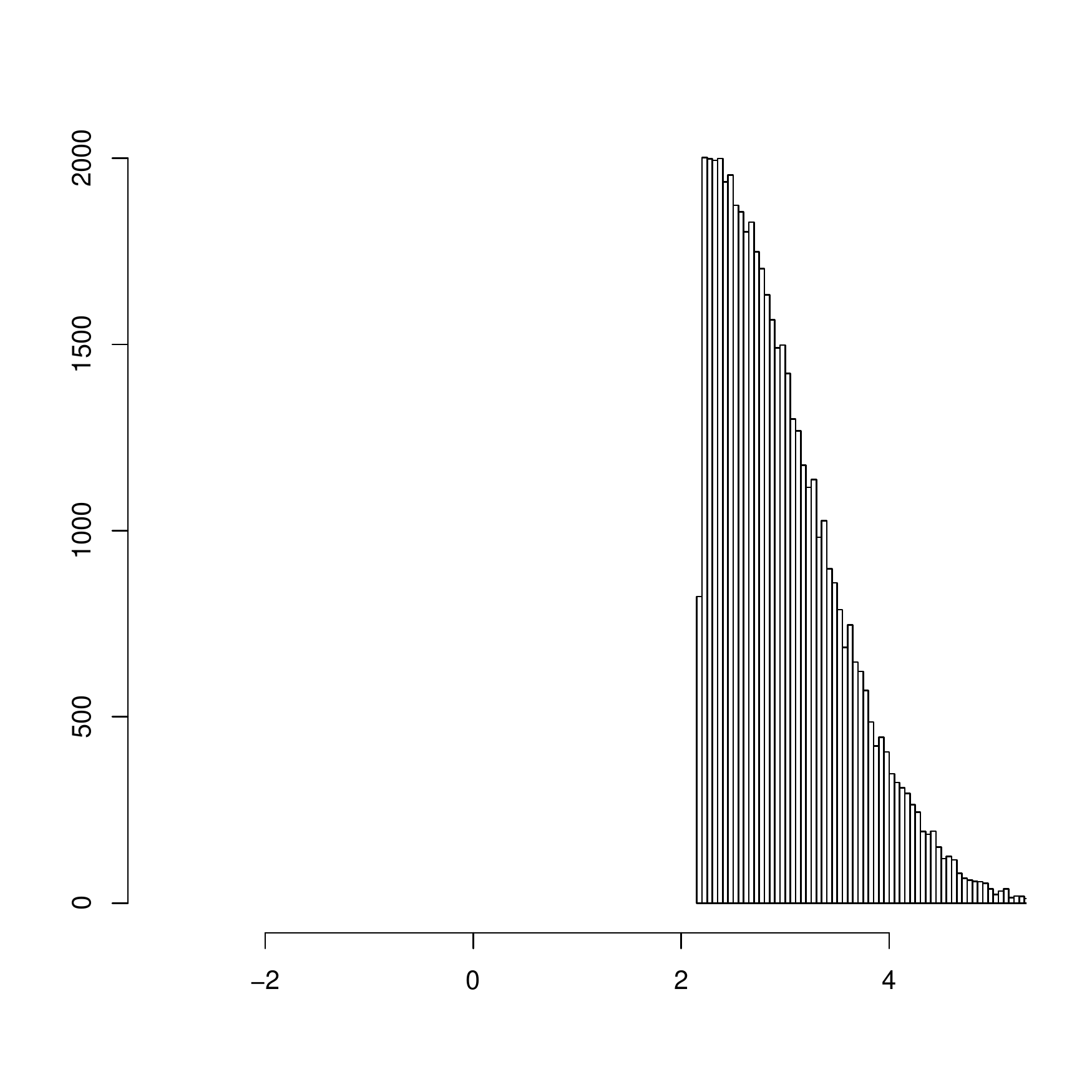}} \quad
\subfigure[$\frac{x_1+x_2}{2}:x_1< 2.18$, $\theta=2.18$]{\includegraphics[width = 1.4in]{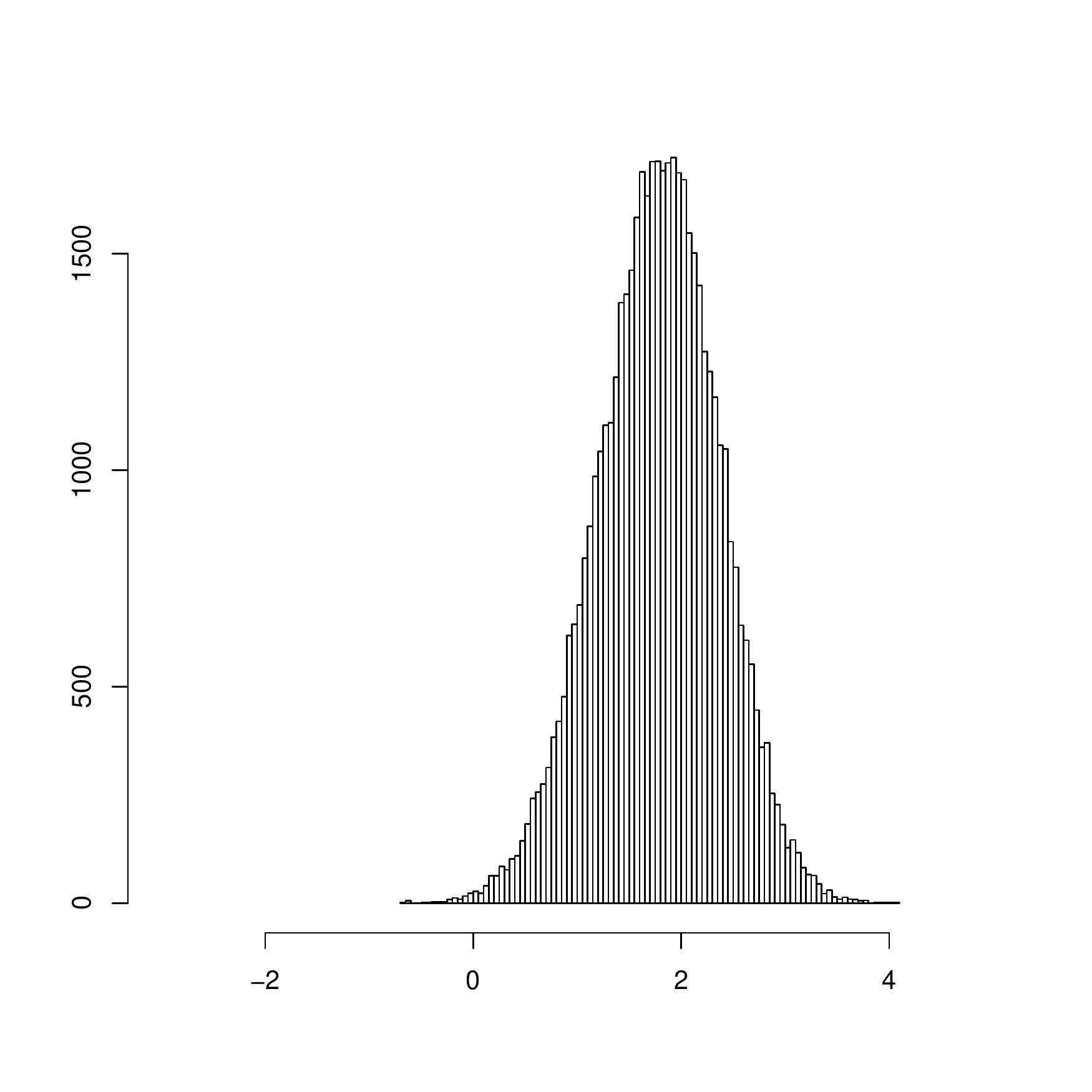}} \quad 
\subfigure[$\widehat\theta$, $\theta=2.18$]{\includegraphics[width = 1.4in]{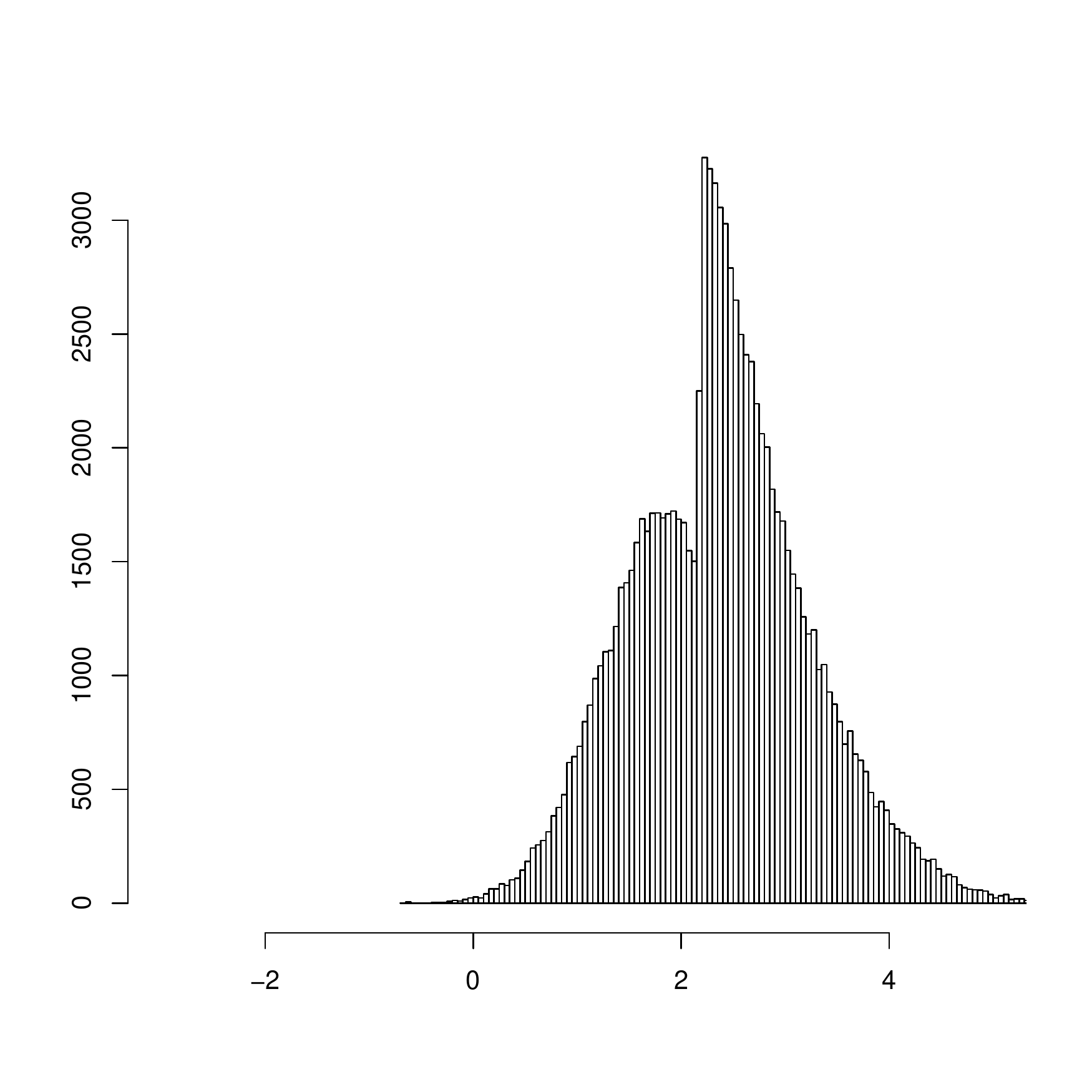}}\quad
\caption{\label{MLEcpic}
Distribution of MLEs from the experiment described in Section~\ref{sec:ex_simple}. $n_1=n_2=1$; $10^6$ Monte-Carlo simulations.}
\label{pic:ex_simple}
\end{figure}

Thus, in this sequential experiment, MLEs are not normal random variables as illustrated in Figure~\ref{pic:ex_simple}. The first column shows the distribution of the MLE if the experiment stopped at stage 1, which is left truncated normal. The seconds of histograms shows the distribution of the MLE if the experiment proceeded to stage 2. This distribution is a mixture of right truncated data from stage 1 and untruncated normal data from stage 2. The final column shows unconditional distribution of the MLE. 

\subsection{The Scope of this Paper}
We consider sequential experiments having a small finite number of interim decision points, that is, experimental set-ups for which Martingale central limit theorems and Brownian theory are not suitable.
Our interest is in experiments that aim primarily on a hypothesis test of effect size. We focus on characterizing the effect of sequential stopping rules on probability distributions of test statistics.

In this manuscript, Section~\ref{sec:densities} introduces notation for GSDs with stopping rules dependent on a parameter of interest [through the distributions of the test statistics]. This section also presents distributions of cumulative test statistics conditional on reaching a stage, conditional on stopping at a stage, and unconditional defined on the probability space with truncation-adapted support. All of these probability distributions are truncated or truncated-mixtures. Section~\ref{Sec:likelihood} presents likelihood-based inference on the truncation-adapted probability space. In section \ref{LSP} a local asymptotic distribution of the MLEs is found to be non-degenerate; it is a mixture of truncated normal distributions. Section \ref{mostpower} shows that the possibility of early stopping does not change the monotonicity of likelihood ratios in the one-parameter exponential family. Thus, stage-specific tests continue to be uniformly most powerful by the Karlin-Rubin theorem, which makes sequential tests based on monotone likelihood ratio uniformly most powerful. Throughout Sections \ref{sec:densities} and \ref{Sec:likelihood} theoretical results are illustrated by a two stage example, \citet{Pocock1977}. Finally, Section \ref{Conclusion} concludes this article with a  summary and a discussion of impact within the contemporary research environment.

\section{Probability Distributions with Early Stopping}\label{sec:densities}
Let  $X$ denote subjects' outcome variable and assume that, when observed in isolation, it has a probability distribution function or a probability mass function $f_X=f_X\left(x|\theta\right)$. To simplify the material, the term \textit{density} is used to refer to probability measures without formally distinguishing between them. Let a sequence of $X$s be observed  with the primary objective of testing the null hypothesis $H_0:\theta=0$ with overall $\alpha$-level type~1 error and $1-\beta$ power at an alternative $H_1: \theta=\theta_1$.

  It is convenient to  group the random sample into \emph{stages} separated by the interim analysis times: $\mathbf{X}_{1}, \mathbf{X}_{2}, \ldots$, where $\mathbf{X}_{k} = (X_{n_{(k-1)}+1},
\ldots,X_{n_{(k)}})$; $n_{(k)}=\sum_{j=1}^kn_j$; here for simplicity  $n_j$ is a pre-specified number of observations in  stage $j$, $1\le  k\le K$;  $n_{(0)}=0$ and $K-1$ is the maximum number of interim analyses permitted. Every stage is assumed to be ``reachable'',  that is, there is a positive probability of reaching each stage.
Data collection at each stage is followed by a hypothesis test that results in a decision to stop the study or to enroll a new group of patients; except that if stage $K$ is reached, the experiment stops after $n_{(K)}$ observations regardless of the last $n_{K}$ observations' values.

\subsection{Stopping Decisions}

Let   $T_{(k)}$ be a function of   observations $\mathbf{X}_{(k)} = (X_1,\ldots,X_{n_{(k)}})$ that is compared against a cutoff value $c_k$  to determine whether to stop at stage $k$  or continue  through stage~$k+1$, $1\le k < K-1$.  These decisions are defined by the events $\left\{\cap_{j=1}^{k-1}\left\{T_{(j)}\le c_j\right\}\right\}\cap \left\{T_{(k)}>c_k\right\}$ and $\cap_{j=1}^{K-1}\left\{T_{(j)}\le c_k\right\}$, respectively, and  are conveniently summarized by a random variable denoting the stopping stage:
$$D = K \cdot I\left(\cap_{j=1}^{K-1}\left\{T_{(j)}\le c_j\right\}\right) 
+ \sum_{k=1}^{K-1} k \cdot I\left(\left(\cap_{j=1}^{k-1}\left\{T_{(j)}\le c_j\right\}\right)\cap \left\{T_{(k)}>c_k\right\}\right);$$
$D\in \{1,\ldots,K\}$ will appear as random index such as  in $\mathbf{X}_{(D)}$ to emphasize that the stopping stage is unknown and is described probabilistically though the random variable $D$. 
$\mathbf{X}_{(d)}$ is the  random variable $\mathbf{X}_{(D)}$ conditioned on stopping with stage $D=d, d=1,\ldots,K$.

It is important to account for $D$ in probability statements about $T_{(d)}$  because $D$ determines  the observations' probability support as illustrated in Section~\ref{sec:ex_simple}.

\subsection{After deciding to stop}
\begin{figure}[!ht]
\centering
\subfigure[Sketch of disjoint support regions for subdensities by  (scaleless) stage-specific test statistics]{\includegraphics[width = 2.5in]{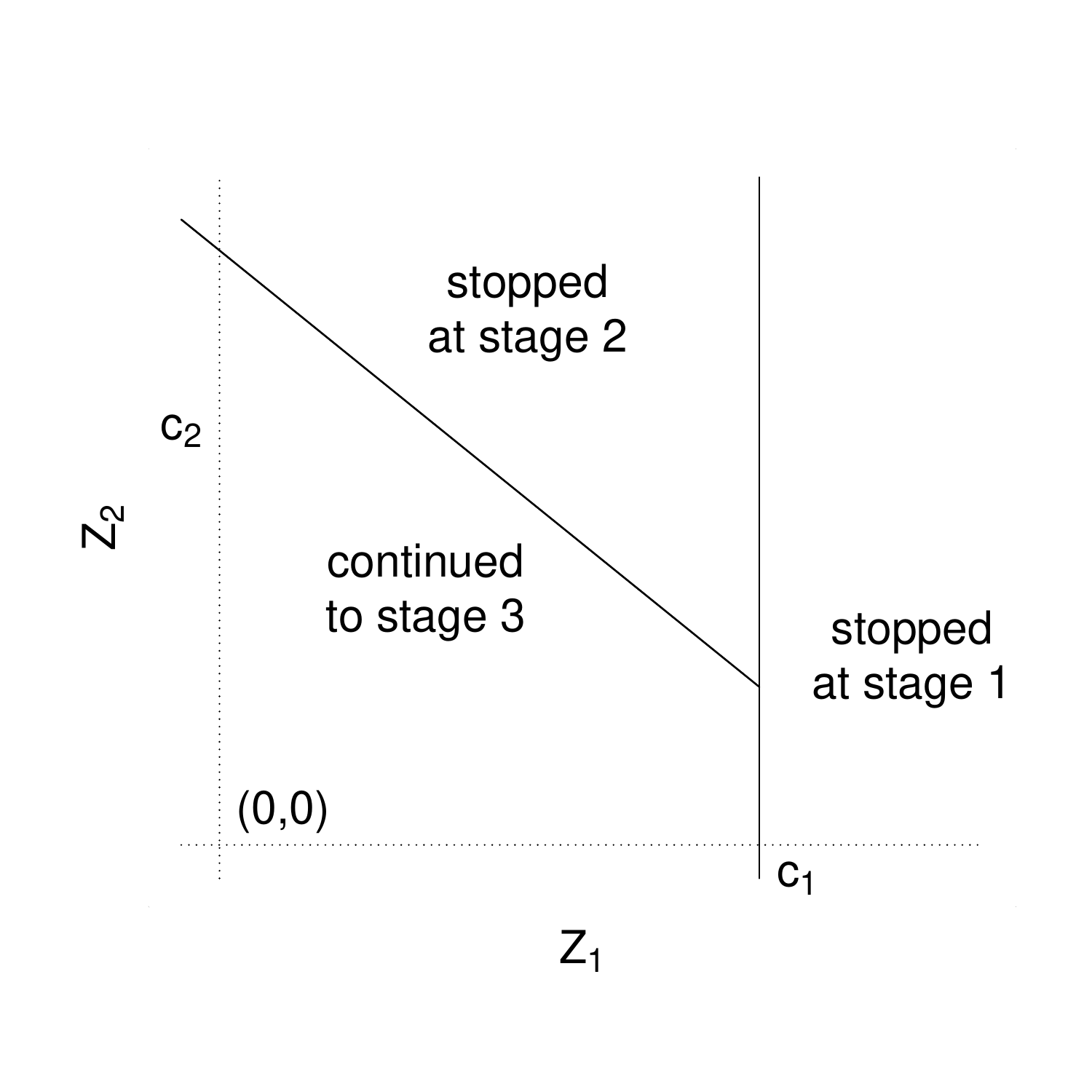}} \quad\quad
\subfigure[Sketch of disjoint support regions for subdensities by  (scaleless) cumulative test statistics]{\includegraphics[width = 2.5in]{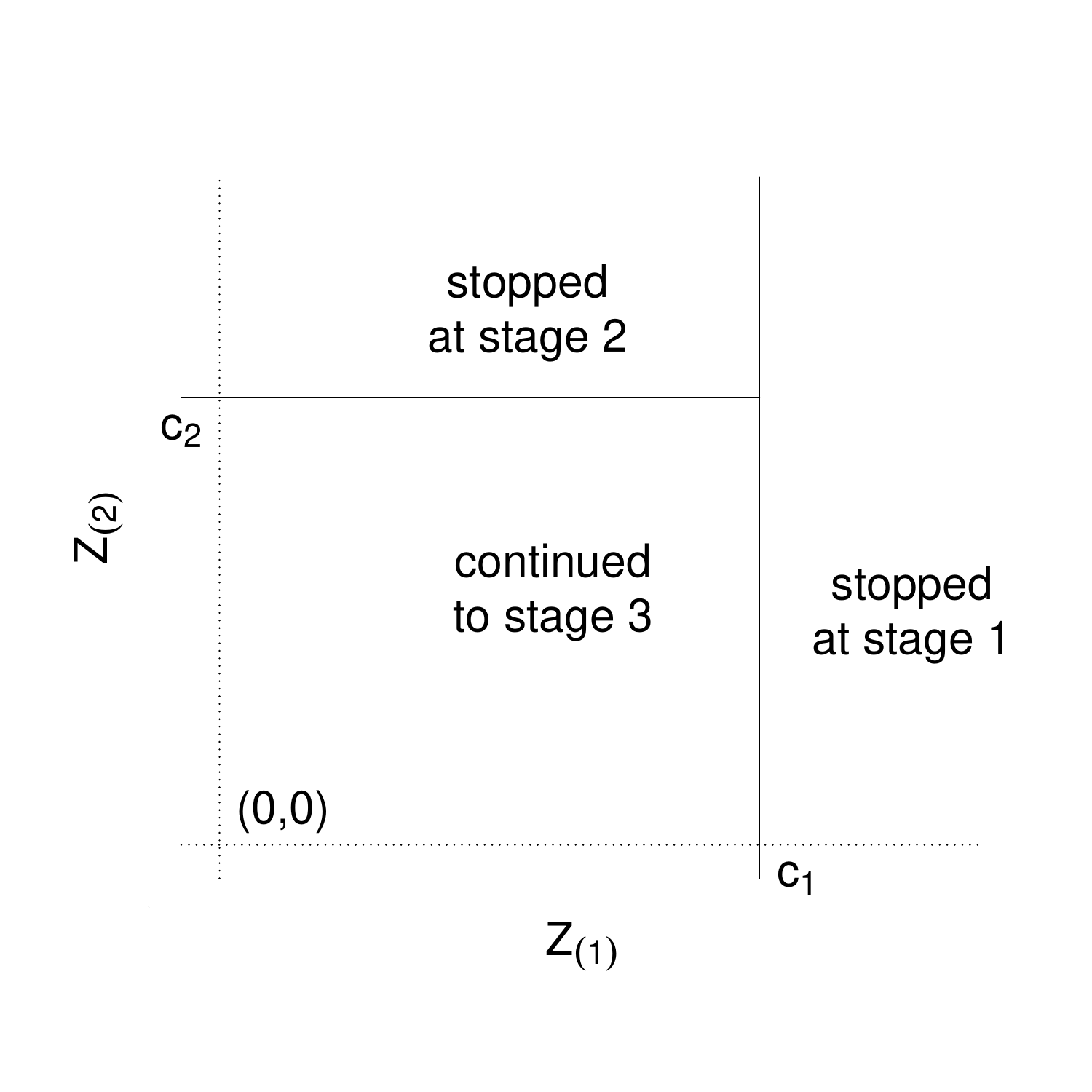}}\\
\caption{\label{support_white}  Support  associated with different stopping decisions when $K=3$;
 $c_k$ is the critical value for stopping at stage $k$.}\label{pic:support_D=3}
\end{figure}
If study stops at stage $D=d<K$, $H_0$ is rejected. If $d=K$, the final hypothesis test determines acceptance or rejection of $H_0$. 
At the time of the final analysis, the density of  observations conditional on stopping at stage $d$ (i.e., the density of  $\mathbf{X}_{(d)}=\mathbf{X}_{(D)}\vert \{D=d\}$) 
 is
\begin{align}\label{densK}
f_{\mathbf{X}_{(d)}}^C=f_{\mathbf{X}_{(D)}} \left(\textbf{\textit{x}}_{(D)}\vert D=d, \theta \right) 
= \frac{I\left(D=d\right)}{\text{Pr}_{\theta}\left(D=d\right)} f_{\mathbf{X}_{(d)}} \left(\textbf{\textit{x}}_{(d)}\vert  \theta \right)
= \frac{f_{\mathbf{X}_{(d)}}^{sub}}{\text{Pr}_{\theta}\left(D=d\right)},\end{align}
where 
$f_{\mathbf{X}_{(d)}}^{sub}=[f_{\mathbf{X}_{(d)}}(\textbf{\textit{x}}_{(d)}|\theta)]^{I\left(D=d\right)}$ denotes the \textit{sub-density} with support defined by $D=d$  (see the exemplary sketches in Figure~\ref{support_white}).
Similarly, for a statistic $T_{(d)} = T(\mathbf{X}_{(d)})$, the conditional on $D=d$ density is $f_{T_{(d)}}^C$.

In contrast, if the stopping rule is not random and the experiment stops with $n_{(d)}$ observations, $\text{Pr}_{\theta}(D=d)\equiv 1$ and the observations have density $f_{\mathbf{X}_{(d)}}$.

\subsection{At the Time of Experimental Design } 

Prior to data collection,  both $D$ and $X_{(D)}$ are unknown and the \textit{joint} density of $X_{(D)}$ can be written in several ways:
\begin{eqnarray} \label{densjoint}
f_{\mathbf{X}_{(D)}}\left(\textbf{\textit{x}}_{(D)}\vert \theta \right) &=&
\sum_{d=1}^{K} [f_{\mathbf{X}_{(d)}}(\textbf{\textit{x}}_{(d)}|\theta)]^{I\left(D=d\right)}\notag \\
&=&\sum_{d=1}^{K}f_{\mathbf{X}_{(d)}}^{sub} = \sum_{d=1}^{K} \text{Pr}_{\theta}\left(D=d\right) f_{\mathbf{X}_{(d)}}^C.\end{eqnarray}
The joint density
is a mixture of densities corresponding to possible outcome vectors; i.e., these densities are defined on non-overlapping regions of the density's support. 

When a test statistic $T_d$ summarizes $d$th stage data, the density of $T_{(D)}$ can be written analogous to  Equation (\ref{densjoint}) as
\begin{eqnarray} \label{dens_cond1}
f_{T_{(D)}}\left({\textit{t}}_{(D)}\vert \theta \right) &=&
\sum_{d=1}^{K}  [f_{{T}_{(d)}}({\textit{t}}_{(d)}|\theta)]^{I\left(D=d\right)}\notag \\
&=&\sum_{d=1}^{K}f_{{T}_{(d)}}^{sub} = \sum_{d=1}^{K} \text{Pr}_{\theta}\left(D=d\right) f_{{T}_{(d)}}^{C}.\end{eqnarray}
Even if every stage-specific test statistic $T_d$ is normally distributed, the distribution of $T_{(D)}$ is not.

\subsection{At  Interim Hypothesis Testing}
Suppose at stage $d-1$, the decision was made to continue sampling,  the support for the $d$th  test statistic is characterized by $D \ge d$.
The density of $T_{(D)}$ conditional on $D \ge d$ is
\begin{eqnarray} \label{densjoint2}
f_{T_{(D)}}\left(t_{(D)}\vert D \ge d, \theta\right) &=&
\frac{\sum_{k=d}^{K}f_{T_{(k)}}^{sub} }{\text{Pr}_{\theta}\left(D\ge d\right)} = \sum_{k=d}^{K} \frac{\text{Pr}_{\theta}\left(D=k\right)}{\text{Pr}_{\theta}\left(D\ge d\right)} f_{T_{(k)}}^{C}.\end{eqnarray}
Again, even if each $T_k$ is normal, the distributions of $T_{(D)}\vert  D \ge d$ are not. 

\subsubsection{Connection with Armitage's algorithm}\label{connectArmitage}
SAS's popular SEQDESIGN procedure, R's gsDesign package, Cytel's EAST and others assess type I and power properties using a recursive sub-density formula (\citet{Armitage1969}) to evaluate the distribution of $T_{(D)}\vert  D \ge d$.  Armitage's subdensity is $f_{T_{(D)}\vert  D \ge d}^{sub} = \sum_{k=d}^K f_{T_{(k)}}^{sub}$ and
\begin{eqnarray} \label{densjoint2_}
f_{T_{(D)}}\left(t_{(D)}\vert D \ge d, \theta\right) &=&
\frac{f_{T_{(D)}\vert  D \ge d}^{sub}}{\text{Pr}_{\theta}\left(D\ge d\right)}.\end{eqnarray}
For example, at $K=2$, the sub-density of $T_{(D)}\vert  D \ge d$ is \begin{equation} \label{arm_subdens}
    f^{sub}_{T_{(D)}\vert  D \ge 2}(t|\theta) = \int_{-\infty}^{c_1} f_{T_{(D\ge 2)}|T_{(1)}} (t|t_1,\theta) f_{T_{(1)}}(t_1|\theta) dt_1
\end{equation} and its density is
$f_{T_{(D)}\vert  D \ge 2}(t|\theta) = f^{sub}_{T_{(D)}\vert  D \ge d}(t|\theta)\left(\int_{-\infty}^{c_1} f_{T_{(1)}}(t_1|\theta)dt_1\right)^{-1}$. 
Recursively, the density conditional on reaching the $d$th interim analysis is
\begin{equation}\label{arm_dens}
    f_{T_{(D)}}(t|D\ge d,\theta) = \frac{\int_{-\infty}^{c_{d-1}} f_{T_{(D \ge d)}|T_{(d-1)}}(t|t_{d-1},\theta) f_{T_{(d-1)}}(t_{d-1}|\theta)  dt_{d-1}}{\int_{-\infty}^{c_{d-1}} f_{T_{(d-1)}}(t_{d-1}|\theta)dt_{d-1}}.
\end{equation}

\subsection{$\sigma-$fields and  support defined by a set of critical values} \label{subspace}

At the time of $d$th hypothesis test, given $D \ge d$, the values  $\{x_j:j > n_{(d)}\}$ are not observable and hence do not contribute to the density; indeed, they  do not belong to the adaptation-rule driven sample space, and consequently, they do not belong to a $\sigma$-field of the random process being monitored: hence, they do not belong to the sequential experiment as a whole. In this paper, by analogy with structural zeroes in contingency tables, these values are excluded from the sample space.

The $d$th stage-specific test statistic $T_{d}=T\left(\textbf{X}_d\right)$ is defined on a probability space $\left({\cal{T}}_d, \sigma\left({\cal{T}}_d\right), P_d\right)$, where ${\cal{T}}_d$ is typically a real line (${\cal{R}}$), $\sigma\left({\cal{T}}_d\right)$ is Borel $\sigma-$field and $P_d$ is a probability measure on the measurable space $\left({\cal{T}}_d,\sigma\left({\cal{T}}_d\right)\right)$. A sequence of nested $\sigma$-fields ${\cal{F}}_{(d)} := \sigma\left({\cal{T}}_1\right) \times \cdots \times \sigma\left({\cal{T}}_d\right),$ $d=1,\ldots,K$, creates a filtration $\textbf{F}=\left({\cal{F}}_{(d)}\right)_{d\le K}$ on the product probability space $\left({\cal{T}}, {\cal{F}}_{(K)}, P\right)$, where ${\cal{T}} = {\cal{T}}_1 \times \cdots \times {\cal{T}}_K$ and $P = P_1 \times \cdots \times P_K$. But in the presence of possible stopping, not all combinations of $(T_1,\ldots,T_K)$ are possible. The cumulative test statistics $T_{(d)} = T_{(d)}(T_1,\ldots,T_d)$ are  defined only on a subspace of the sample space ${\cal{T}}_1 \times \cdots \times {\cal{T}}_d$. Thus, probability environment substantially changes.

\subsubsection{Interim hypothesis testing}
The  statistic $T_{(1)}=T_1$ conditional on reaching stage $1$ $(D\ge1)$ is defined on the sample space ${\cal{T}}_{(1)} = {\cal{T}}_1$; so
 $\sigma\left({\cal{T}}_{(1)}\right) = {\cal{F}}_{(1)}$. The statistic $T_{(2)}$ conditional on $D\ge 2$ is defined on $${\cal{T}}_{(2)} = \left[(t_1,t_2): \{t_1 \le c_1\}\right],$$
and $\sigma\left({\cal{T}}_{(2)}\right) \subset {\cal{F}}_{(2)}$. Further, for $d\in\{3,\ldots,K\}$, $T_{(d)}$ conditional on $D\ge d$ is defined on $$
{\cal{T}}_{(d)} = \left((t_1,\ldots,t_d): \left\{t_{(j)} \le c_j \right\}, j=1,\ldots,d-1 \right).
$$ 
For all $d$, the support ${\cal{T}}_{(d)}$ and the $\sigma$-field $\sigma\left({\cal{T}}_{(d)}\right)$ is reduced by the possibility of early stopping: $\sigma\left({\cal{T}}_{(d)}\right) \subset {\cal{F}}_{(d)}$. This creates new measurable spaces $\left({\cal{T}}_{(d)}, \sigma\left({\cal{T}}_{(d)}\right)\right)$ for interim tests at every  possible stage $1,\ldots,K$. Armitage's recursive sub-density formula [\citet{Armitage1969}] is defined on this measurable space; see Equation (\ref{arm_subdens}) for $K=2$. Re-scaling yields a density function [see Equation (\ref{arm_dens})] which defines a probability measure to complete the probability space $\left({\cal{T}}_{(d)}, \sigma\left({\cal{T}}_{(d)}\right), P_{(d)}\right)$, where the probability measure  $P_{(d)}$ is  determined by density (\ref{arm_dens}): $P_{(d)}(A) =  \int_A f_{T_{(D)}}(t|D\ge d,\theta) dt$, $A \in  \sigma\left({\cal{T}}_{(d)}\right)$.

\subsubsection{After the  stop decision is made}
At each interim stage $d=1,\ldots,K-1$, the decision to reject or accept $H_0$ splits ${\cal{T}}_{(d)}$ into two non-overlapping regions denoted ${\cal{T}}_{(d)}^{stop}$ and ${\cal{T}}_{(d)}^{cont}$, respectively. Since 
${\cal{T}}_{(d)}^{stop} \cap {\cal{T}}_{(d+1)} = \emptyset$ and
${\cal{T}}_{(d)}^{cont} \subset {\cal{T}}_{(d+1)}$,  then the sets ${\cal{T}}_{(1)}^{stop},\ldots,{\cal{T}}_{(K-1)}^{stop},$ and ${\cal{T}}_{(K)}$ make a partition of the sample space of $T_{(D)}$:

$${\cal{T}}_{(K)} + \sum_{k=1}^{K-1}{\cal{T}}_{(k)}^{stop} =  \left(\cup_{d=1}^{K-1}{\cal{T}}_{(d)}^{stop}\right) \cup {\cal{T}}_{(K)} = \cup_{d=1}^K{\cal{T}}_{(d)} \subset {\cal{T}}.$$ 
We define a probability space for the observable random variable ${T}_{(d)}$ using the probability measures $P_{(d)}$ defined by $f_{T_{(d)}}^C$ on the measurable space $\left({\cal{T}}_{(d)}^{stop}, \sigma\left({\cal{T}}_{(d)}^{stop}\right)\right)$.
At $D=K$, the probability measure $P_{(K)}$ defined by $f_{T_{(K)}}^C$ completes the probability space for a measurable space $\left({\cal{T}}_{(K)}, \sigma\left({\cal{T}}_{(K)}\right)\right)$.

Thus, at the end of the study, at $D=d$, the researcher operates with an \emph{observable} random variable $T_{(d)}$.

\subsubsection{At the design stage} The conditional random variables, $T_{(d)}$, defined on non-overlapping $\sigma$-fields are combined together into the unconditional random variable $T_{(D)} \sim f_{T_{(D)}}$ defined on the sample space $\left(\cup_{d=1}^{K-1}{\cal{T}}_{(d)}^{stop}\right) \cup {\cal{T}}_{(K)}$. The probability distribution defined on the $\sigma$-field on this sample space is $P_{(D)} = \sum_{d=1}^K\text{Pr}\left(D=d\right) P_{(d)}$.

\subsubsection{Impossible events}
The set ${\cal{T}}_0 = {\cal{T}} \setminus \cup_{d=1}^K{\cal{T}}_{(d)}$ contains all  impossible combinations of $(t_1,\ldots,t_K) $ under a chosen stopping rule.  If $K=2$, for example, then ${\cal{T}} \setminus \cup_{d=1}^2{\cal{T}}_{(d)} = \left\{(t_1,t_2): t_1>c_1\right\}$. Without the possibility of early stopping, all combinations of $(t_1,\ldots,t_K)\in{\cal{T}}$ would be possible.

\subsection{Pocock's Example: One-Sided Two-Group Sequential Z-test}\label{Pocock_example}

\citet{Pocock1977} proposed a simple two-stage procedure for testing $H_0: \theta = 0$ with a pre-determined power at $H_1: \theta = \theta_1$ on normal data. With $n_1=n_2$, $c_1=c_2=2.18$ is used to secure the overall type~1 error rate $\alpha=0.025$ with a one-sided $z$ test. Let $n_0=0$ and
$$Z_k = \frac{1}{\sqrt{n_{k}}}\sum_{i=n_{(k-1)}+1}^{n_{(k)}} X_{i} = \sqrt{n_{k}} \cdot \bar X_k  \overset{d}{=} \mathcal{N}(\theta,1),\quad k=1,2.$$ 
The study is stopped for efficacy at stage 1 if $Z_1\ge 2.18$ and proceeds to stage 2 when $Z_1 < 2.18$. If the study is stopped at stage~1,  the support for $\bar X_1|Z_1\ge 2.18$ starts at $2.18/\sqrt{n_1} $ and stretches to $+\infty$. If the study continues through stage~2, support for $\bar X_1|Z_1<2.18$  ranges from $-\infty$ to $2.18/\sqrt{n_1}$ . 

The test statistic for the second stage, under $Z_1<2.18$,
$$Z_{(2)}=\frac{\sqrt{n_1}}{\sqrt{n_1+n_2}}Z_1 + \frac{\sqrt{n_2}}{\sqrt{n_1+n_2}}Z_2 =  \frac{1}{\sqrt{n_1+n_2}}\sum_{i=1}^{n_1+n_2}X_i.$$
The distribution of $Z_{(2)}$ is a mixture of a right-truncated normal $Z_1 |Z_1 < 2.18$
and the normal $Z_2$.

\begin{figure}[htb!]
\centering
\subfigure[$Z_1 |Z_1 \ge 2.18$, $\theta=2.18$]{\includegraphics[width = 1.7in,height=.18\textheight]{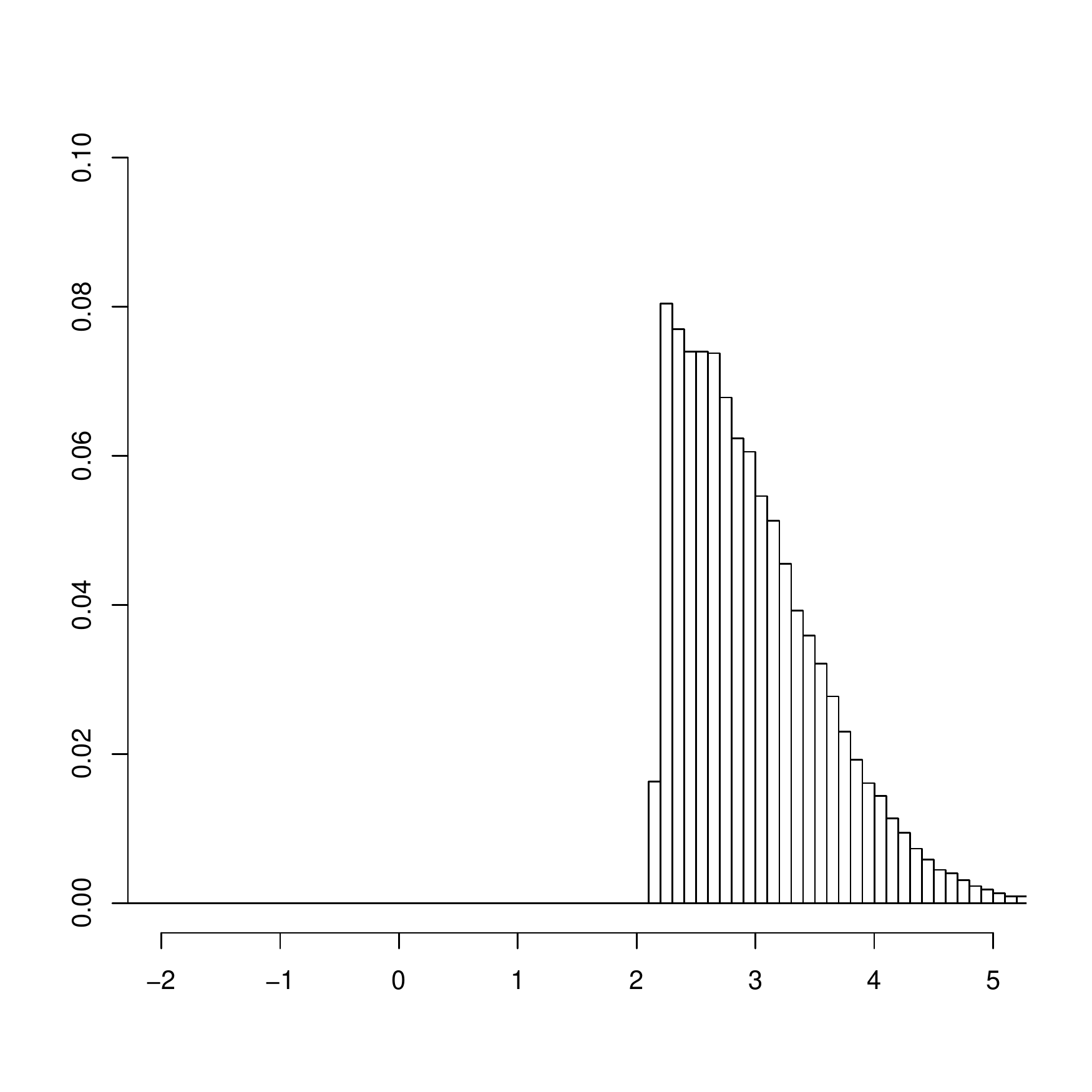}} 
\subfigure[$Z_{(2)} |Z_1 < 2.18,$ $\theta=2.18$]{\includegraphics[width = 1.7in,height=.18\textheight]{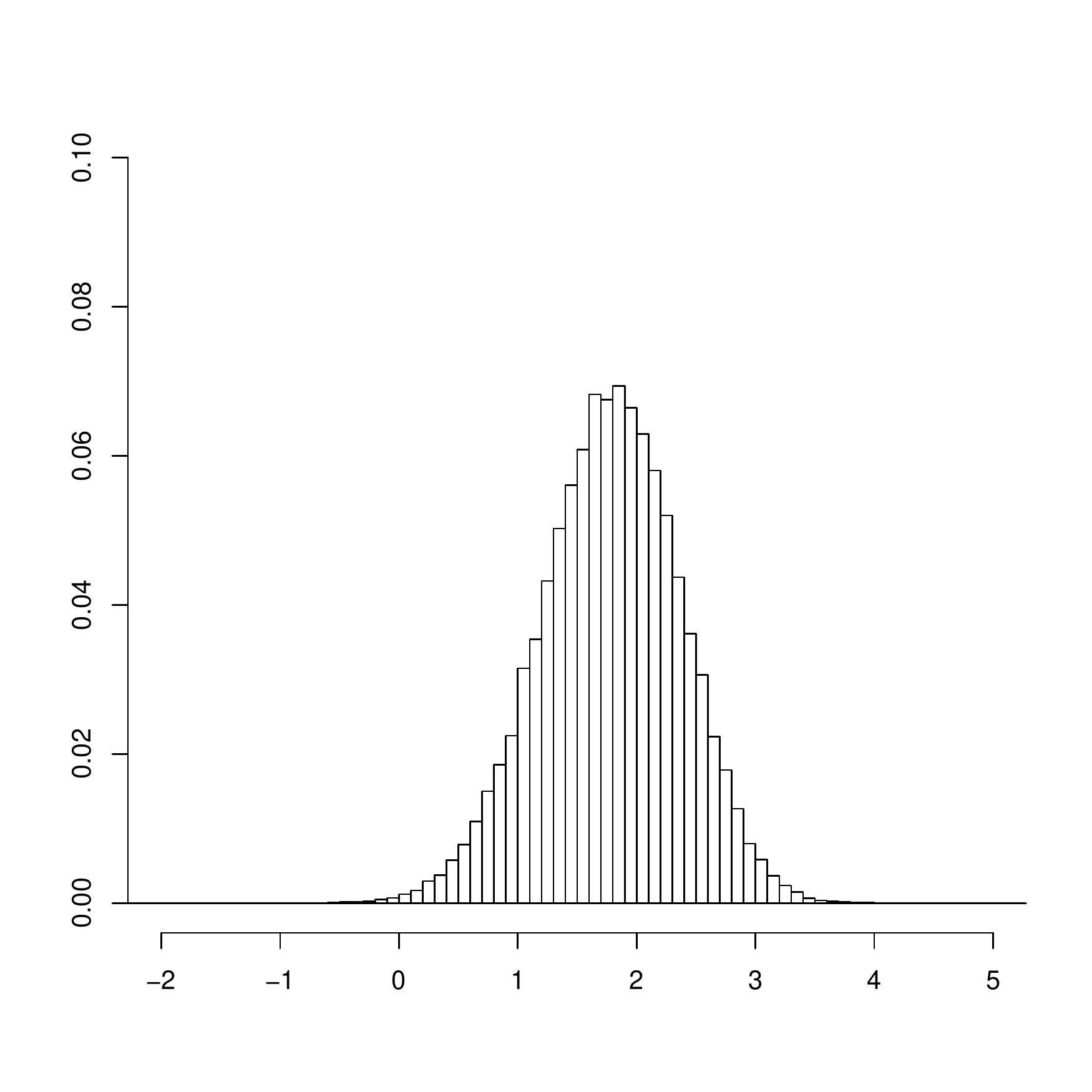}} 
\subfigure[$Z_{(D)},\theta=2.18$]{\includegraphics[width = 1.7in,height=.18\textheight]{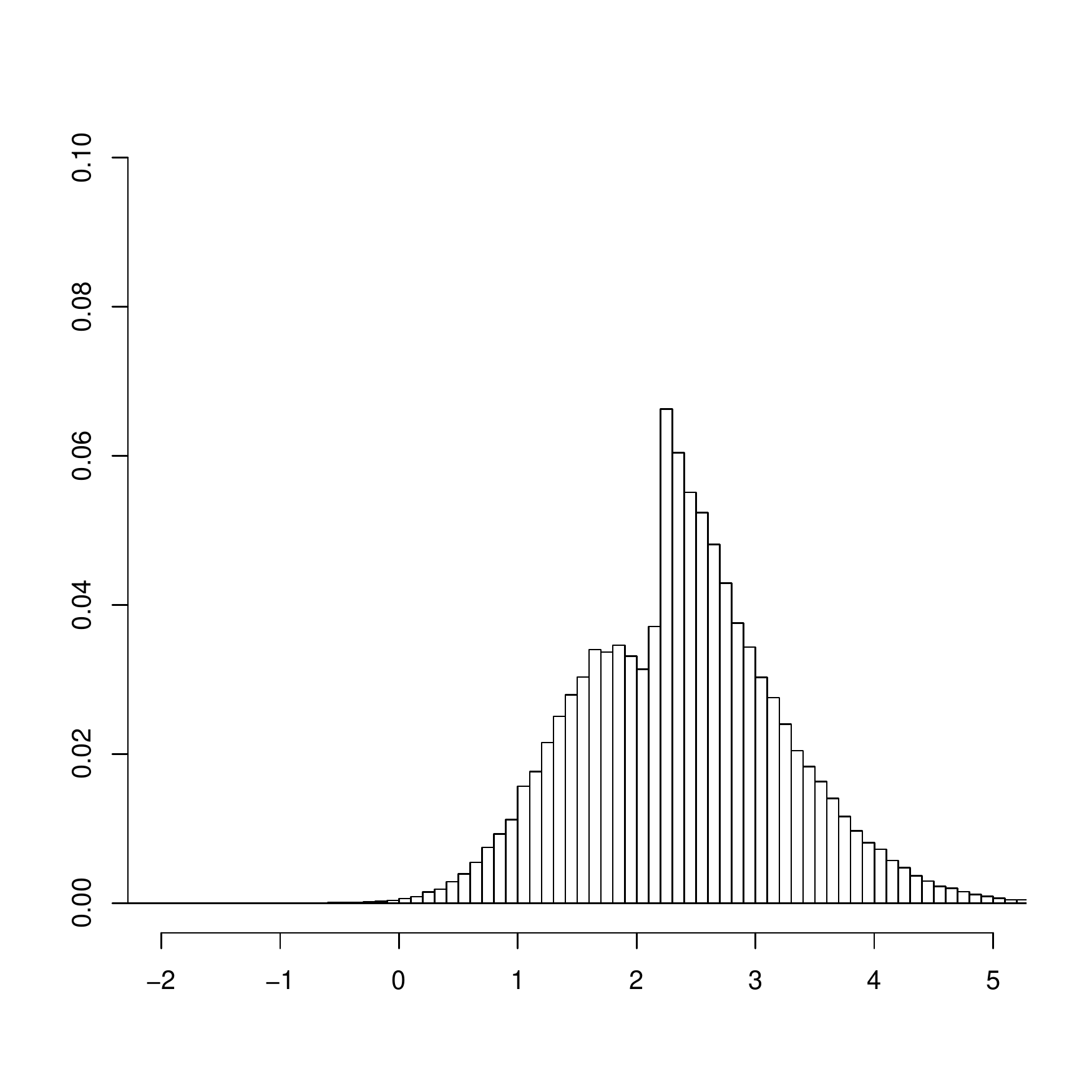}} 
\caption{\label{Z_example} Histograms of test statistics from a two-stage Pocock's experiment with critical value  $c_1=2.18$; $f_{\mathbf{X}_k}=\mathcal{N}(2.18,1)$ and $n_k=100, k=1,2$.}
\end{figure}

Figure~\ref{Z_example} shows histograms of $Z_1 |Z_1 \ge 2.18$, $Z_{(2)} |Z_1 < 2.18$, and
$Z_{(D)}$ 
estimated from $100,000$ Monte-Carlo samples assuming $\theta=2.18$. These histograms based on $n_1=n_2=100$ are almost identical  to the histograms in Figure \ref{MLEcpic} based on $n_1=n_2=1$. This highlights on the important message that non-normality continues to be present even asymptotically \cite{Tarima2019}. 

\section{Likelihood-based Inference with Early Stopping } \label{Sec:likelihood}
If the stopping rule  is not random and the experiment stopped with $n_{(d)}$ observations, $\text{Pr}_{\theta}(D=d)\equiv 1$, the likelihood is
\begin{eqnarray} \label{Lfix}
{\mathcal{L}}^{fix}\left(\theta \vert k,\textbf{\textit{x}}_{(d)}  \right) =
f_{\mathbf{X}_{(d)}}.\end{eqnarray}
Considering the joint density \eqref{densjoint}  conditional on the  data  $\left(d,\textbf{\textit{x}}_{(d)}\right)$ observed at the end of a sequential experiment, the   likelihood is
\begin{eqnarray} \label{LKjoint}
{\mathcal{L}}\left(\theta \vert d,\textbf{\textit{x}}_{(d)} \right) =
f_{\mathbf{X}_{(d)}}^{sub}.\end{eqnarray} 
%
 
The indicator  in \eqref{densK} emphasizes that support for the random variables is reduced by the conditioning; see this illustrated in Figure~\ref{support_white}.
Note also in Figure~\ref{support_white} that the support conditional on stopping at one stage is disjoint from the support conditional on stopping at another stage.

${\cal{L}}$ is a function of $\theta$ and  the observed data $(d, \textbf{\textit{x}}_{(d)})$, but ${\cal{L}}$ is a continuous function only of $\theta$, as discontinuities in $(d, \textbf{\textit{x}}_{(d)})$ arise from  the mixture distribution of $\mathbf{X}_{(D)}$. These discontinuities are inherited by MLEs and other statistics derived from them.
Conditional on $(d, \textbf{\textit{x}}_{(d)})$, MLEs maximizing ${\mathcal{L}}$ 
are
 \begin{align*}
\widehat\theta &= \arg \max_{\theta} f_{\mathbf{X}_{(d)}}^{sub}. 
\end{align*}
For every  $\left(d,\textbf{\textit{x}}_{(d)}\right)$, $f_{\mathbf{X}_{(d)}}^{sub} =f_{\mathbf{X}_{(d)}}$ for all $\theta$;  consequently, 
$$\widehat\theta^{fix} := \arg \max_{\theta} f_{\mathbf{X}_{(d)}} = \arg \max_{\theta} f_{\mathbf{X}_{(d)}}^{sub} = \widehat\theta,$$ 
that is, as is well known, making stopping decisions does not alter maximum likelihood point estimates; they are  the same whether obtained by maximizing ${\mathcal{L}}$ or ${\mathcal{L}}^{fix}$ and
 other observed statistics derived from the likelihood (e.g., the score function and the observed information) are unaffected as well.

The following example extends the simple one in Section~\ref{sec:ex_simple} with $n_1=n_2=1$ to arbitrary $n_1$ and $n_2$ to illustrate (as is proven later) that, although maximum likelihood point estimates and test statistics are unaffected by early stopping decisions,  their probability distribution does not tend to normality even with larger sample sizes.

\subsection{Large Sample Properties}\label{LSP} 
The $d$th stage-specific MLEs $\widehat\theta_d$ of $\theta$ and their statistical models are called 
\emph{regular} if, without the possibility of early stopping,
\begin{equation} \label{ctlholds}
\xi_d= \sqrt{n_d}\left(\widehat\theta_d - \theta\right)
\overset{d}{\to} {\cal{N}}(0,\sigma^2),
\end{equation}
where $0<\sigma < \infty$. Assumption \eqref{ctlholds} was
described  in \citet{Tarima2019} to  include the more 
specific assumptions for
\begin{itemize} 
\item  independent, identically distributed observations by Cram\'er [e.g., for example, \citet{Ferguson1996}], 
\item  independent not identically distributed observations [e.g., \citet{philippou1973asymptotic}],
\item  dependent observations  [e.g., \citet{crowder1976maximum}], 
\item and  densities whose support depends on parameters [e.g., \citet{Wang14}]. 
\end{itemize}
All these specific sets of assumptions include assumptions of the existence and consistency of the MLE. 

With large samples, the MLE at the stage  $d$ analysis can be approximated recursively by
\begin{align}
\widehat\theta_{(d)} &\approx \frac{n_{(d-1)}}{n_{(d)}}\widehat\theta_{(d-1)} + 
                 \frac{n_d}{n_{(d)}}\widehat\theta_d \label{MLE_d} \approx \sum_{j=1}^d \frac{n_j}{n_{(d)}} \widehat\theta_d,
\end{align}\label{eq:MLE}
where  $\widehat\theta_{(d-1)}$ is an MLE based on cumulative data from stages $1$ to $d-1$ and $\widehat\theta_{d}$ is the $d$th stage-specific MLE. The standardized $d$th stage-specific  MLE is so, 
$$T_{(d)}
= \sqrt{n_{(d)}} \sum_{j=1}^d \frac{n_j}{n_{(d)}} \left(\frac{\widehat\theta_j-\theta}{\sigma}\right) =
 \sum_{j=1}^d 
 \sqrt{\frac{n_j}{n_{(d)}}}\ \xi_j \approx  \sqrt{n_{(d)}}  \left(\frac{\widehat\theta_{(d)}-\theta}{\sigma}\right),$$
that is, $T_{(d)}$ is (approximately) the standardized $d$th stage-specific MLE.  
The asymptotic properties of $T_{(D)}$ 
 depend on the existence and distribution of a limiting random variable $r_{(D)}$  defined by
\begin{equation}\label{tau_n}
\sum_{d=1}^K I(D=d)\frac{n_{(d)}}{n_{d}} \overset{d}{\to} \sum_{d=1}^K I(D=d)r_{(d)} = r_{(D)},
\end{equation}
where $r_{(d)}=\lim_{n_{d}\to\infty} n_{(d)}/n_{d}$ is the asymptotic ratio of the $d$th cumulative-stage and  stage-specific sample sizes
 [Theorem 1 in \citet{Tarima2019}]; 
 $r_{(D)}$ is a multinomial random variable with support on $r_{(d)}$. 

Assume the limits $n_{(d)}/n_j \to r_{(d)j} \in (0,\infty)$, $j\le d$ exist with $r_{(d)d} = r_{(d)}$.  Then given   $D = d$,
$$T_{(d)} 
\to \sum_{j=1}^d \frac{\xi_j}{\sqrt{r_{(d)j}}}.$$
While $\xi_j \overset{d}{\to} {\cal{N}}(0,1)$, 
the distributions used for the final analysis, the interim analysis and experimental design, respectively, are
mixtures of truncated distributions:
\begin{eqnarray*}{\text{Pr}_{\theta}}\left(T_{(D)} < v|D=d \right) &\to& 
\text{Pr}_{\theta}\left(\sum_{j=1}^d \frac{\xi_j}{\sqrt{r_{(d)j}}}  < v \Big\vert D=d \right),\\ 
{\text{Pr}_{\theta}}\left(T_{(D)} < v | D\ge d\right) &\to& 
\sum_{k=d}^K {\text{Pr}_{\theta}}\left(D=k\right){\text{Pr}_{\theta}}\left(\sum_{j=1}^k \frac{\xi_j}{\sqrt{r_{(k)j}}}  < v \Big\vert D=k \right),\\
\text{Pr}_{\theta}\left(T_{(D)} < v \right) &\to& 
\sum_{k=1}^K {\text{Pr}_{\theta}}\left(D=k\right){\text{Pr}_{\theta}}\left(\sum_{j=1}^k \frac{\xi_j}{\sqrt{r_{(k)j}}}  < v \Big\vert D=k \right).
\end{eqnarray*}

\subsection*{Pocock's Example: Large Sample Properties}\label{sec:example}
Under assumption \eqref{ctlholds} with large sample sizes,     
$\widehat\theta_{(D)} = I(D=1)\widehat\theta_{(1)} + I(D=2)\widehat\theta_{(2)}$
are standardized as
\begin{align}
T_{(D)} &= \sqrt{n_{(D)}}\left(\widehat\theta_{(D)} - \theta\right)/\sigma \notag \\
&= I(D=1)\sqrt{n_1}\left(\widehat\theta_{(1)} - \theta\right)/\sigma + 
I(D=2)\sqrt{n_1+n_2}\left(\widehat\theta_{(2)} - \theta\right)/\sigma.\end{align}
Assume the limit 
$I(D=1) + I(D=2)\frac{n_1+n_2}{n_2} \overset{d}{\to} r_{(D)}$ exists  as  $n_1\to\infty$,
   Then, adopting the local 
 alternative hypothesis $\theta = h / \sqrt{n_1}$ yields a non-degenerate limiting distribution of $T_{(D)}$ that models both stages of the experiment:
\begin{eqnarray}\text{Pr}_{\theta}\left(T_{(D)} < v \right) &\to& p_1 \Phi\left(\left. v\right\vert D=1\right) \notag \\
&+& (1-p_1) \int_{-\infty}^{c_1} \Phi\left(
\sqrt{r_{(D)}} v - \sqrt{r_{(D)}-1} y\right) \phi\left(y|D=2\right) dy,
\end{eqnarray} where $p_1 = \lim_{n_1\to\infty}{\text{Pr}_{\theta}}\left(D=1\right)$ is the limiting stage 1 stopping probability; $\phi$ and $\Phi$ denote the standard normal density and cumulative distribution function, respectively.  With a fixed alternative $\theta$, 
$\text{Pr}_{\theta}\left(T_{(D)} < v \right) \to \Phi(v)$ because $p_1\to 1$ with probability 1, and modeling related to stage~2 data is lost.

In Pocock's example, 
$T_{(D)} = I(D=1)Z_1 + I(D=2)\left(Z_1 + Z_2\right)/\sqrt{2}$ and 
\begin{eqnarray}
\text{Pr}_{\theta}\left(T_{(D)} < v \right) &\to& \text{Pr}_{\theta}\left(Z_1>c_1\right) \Phi\left(\left. v\right\vert Z_1>c_1 \right) \notag \\
&+& \text{Pr}_{\theta}\left(Z_1\le c_1\right) \int_{-\infty}^{c_1} \Phi\left(
\sqrt{2} v - y\right) \frac{\phi\left(y\right)}{\text{Pr}\left(y \le c_1\right)}dy
\end{eqnarray}
Note if $Z_1\le c_1$, then $$\text{Pr}_{\theta}\left(\frac{Z_1+Z_2}{\sqrt{2}} < v \vert Z_1\le c_1\right) = 
\int_{-\infty}^{c_1} \Phi\left(
\sqrt{2} v - y\right) \frac{\phi\left(y\right)}{\text{Pr}\left(y \le c_1\right)} dy,$$
which is a continuous mixture of distributions.

\subsection{Most Powerful Group Sequential Tests} \label{mostpower}
Let $X\sim f_X(\theta)$, where $f$ belongs to the one-parameter exponential family. Without a possibility of early stopping, the likelihood for a realization $\textbf{\textit{x}}=\left(x_1,\ldots,x_n\right)$ of a random sample $\textbf{X}=\left(X_1,\ldots,X_n\right)$ is
\begin{align*}
{\cal{L}}\left(\theta|\textbf{\textit{x}}\right) = \prod_{i=1}^n f_X\left(\textbf{\textit{x}}\right) &= 
h\left(\textbf{\textit{x}}\right) g \left(T\left(\textbf{\textit{x}}\right) |\theta \right) = 
h\left(\textbf{\textit{x}}\right)  e^{\eta(\theta)T\left(\textbf{\textit{x}}\right) + A\left(\theta\right)}
\end{align*}
where all relevant information about $\theta$ is absorbed by a sufficient statistic $T\left(\textbf{\textit{x}}\right)$.  Assume the test statistic $Z$ is a one-to-one transformation of $T$.

If $LR(t)=g(t|\theta)/g(t|\theta_0)$ has monotone likelihood ratio (MLR) in $t$, then the Karlin-Rubin theorem provides uniformly most powerful (UMP) tests.

In sequential testing settings, when $d$th stage is reached $\left(D\ge d\right)$, the interim likelihood \begin{eqnarray}
{\cal{L}}\left(\theta|D\ge d,\textbf{\textit{x}}_{(d)}\right)
&=& I\left(D\ge d\right) h\left(\textbf{\textit{x}}_{(d)}\right) e^{\eta(\theta)T\left(\textbf{\textit{x}}_{(d)}\right) + A_d\left(\theta\right)}
\label{Lu}
\end{eqnarray} 
has the associated interim likelihood ratio
\begin{eqnarray}
LR(t|D\ge d) = \frac{{\cal{L}}\left(\theta|D\ge d,\textbf{\textit{x}}_{(d)}\right)}{{\cal{L}}\left(\theta_0|D\ge d,\textbf{\textit{x}}_{(d)}\right)}.
\notag
\end{eqnarray}
For every $d$th stage hypothesis test, $D\ge d$ and $$
LR(t|D\ge d)= \exp \left[\left(\eta(\theta_1)-\eta(\theta_0)\right) T_{(d)} + \left(A_d(\theta_1)-A_d(\theta_0)\right)\right]$$ which means that \textit{the MLR property is 
preserved with early stopping}.

\textbf{Definition:} A sequence of $\alpha_d$-level interim tests $\{T_{(1)} > c_1\},\ldots,\{T_{(D)} > c_D\}$ will be called the sequential test.

\begin{theorem} \label{theorem1} For any fixed $(\alpha_1,\ldots,\alpha_K)$ and $\left(n_1,\ldots,n_K\right)$ $(1)$ the interim LR test $\{T_{(d)} > c_d\}$ is a UMP $\alpha_d$-level test and $(2)$ no sequential test is more powerful than the  sequential test based interim LRs.
\end{theorem}
\begin{table}[bt!]
	\centering
	\begin{tabular}{|c|c|}
		\hline
		Feature & Mathematical Definition \\
		\hline
		\textit{overall type $1$ error} & $\alpha = \sum_{k=1}^K \alpha_k \prod_{j=1}^{k-1}\text{Pr}_{0}\left(Z_{(j)}\le u_{j}\right)$\\
	\textit{stage-specific type 1 error} & $\alpha_d = \text{Pr}_{0}\left( Z_{(d)} > u_d \big| \cap_{j=1}^{d-1} \left\{Z_{(j)}\le u_{j} \right\}\right)$ \\
	\textit{$\alpha$-spending function} & $\alpha_{(d)} = \sum_{j=1}^d\alpha_j \prod_{i=1}^{j-1}\text{Pr}_{0}\left(Z_{(j)}\le u_{j}\right)$\\
    \hline
		\textit{overall type 2 error} & $\beta(\theta)=\sum_{j=1}^K\beta_j(\theta) \prod_{i=1}^{j-1}\text{Pr}_{\theta}\left( Z_{(j)}\le u_{j}\right)$\\
	\textit{stage-specific type 2 error} & $\beta_d(\theta) = \text{Pr}_{\theta}\left( Z_{(d)} > u_d \big| \cap_{j=1}^{d-1} \left\{Z_{(j)}\le u_{j} \right\} \right)$\\
	\textit{$\beta$-spending function} & $\beta_{(d)}\left(\theta\right) = \sum_{j=1}^d\beta_j(\theta) \prod_{i=1}^{j-1}\text{Pr}_{\theta}\left(Z_{(j)}\le u_{j}\right)$\\
	\hline
		\textit{overall power} & $1-\beta(\theta)$\\
	\textit{stage-specific power} & $  1-\beta_d(\theta)$\\
	\textit{cumulative power} & $1-\beta_{(d)}\left(\theta\right)$\\
		\hline		\end{tabular}
	\caption{Definitions of Operational Characteristics for  Sequential Tests $d=1,\ldots,K$; $\prod_{i=1}^0[\cdot] = 1$.}
	\label{Definitions}
\end{table}
\textbf{Proof.} 
$(1)$ As shown above,  conditional on reaching stage $d$, $(D\ge d)$, $T_{(d)}$ continues to be sufficient  in exponential families and the MLR property is  preserved. By the Karlin-Rubin theorem, the test based on $T_{(d)}$ is uniformly most powerful at $d^{th}$ stage. \\
$(2)$ 
Using Table \ref{Definitions} notation, 
$$1-\beta(\theta) 
= \sum_{k=1}^K \left(\prod_{j=1}^{k-1}\left[\beta_j(\theta)\right] \right) [1-\beta_k(\theta)]= 1 - \prod_{k=1}^{K}\beta_k(\theta).$$

At $K=1$, $\alpha_1$ and $n_1$ uniquely define $c_1$ and $\left\{T_{(1)} > c_{1}\right\}$ is a UMP by part 1 with a stage-specific power curve $1-\beta_1(\theta)$. 
At an arbitrary stage $d$, $\alpha_d$ and $n_d$ uniquely define $c_d$ and, by part 1 of the theorem, the stage-specific power $1-\beta_d(\theta)$, is the highest. Thus, for any choice of $(\alpha_1,\ldots,\alpha_K)$ and $\left(n_1,\ldots,n_K\right)$, and consequently $\{T_{(d)} > c_d\}$, the power of the sequential test based on LRs is $1-\beta(\theta) = 1-\prod_{k=1}^K\beta_k(\theta)$. This power is the highest for any given $\theta$, because the stage specific type 2 errors $\beta_d(\theta)$ are the lowest for each $d$ at any $\theta$. 
\textbf{Q.E.D.}

\subsection*{Pocock's Example: Likelihood Ratio} For $X_i\sim {\cal{N}}(\theta,1)$, $i=1,\ldots,n$, the likelihood ratio for testing  $H_0:\theta=\theta_0$ vs $H_A:\theta=\theta_1>\theta_0$ is 
$$\log LR(t) = -n\left(\log \theta_0 - \log \theta_1\right) - t/\theta_1,$$
which has rejection region $\left\{t>c\right\}$, where $t = \sum_{i=1}^n X_i$ is a sufficient statistic. 
Using $Z=t/\sqrt{n} \sim f_{Z} = {\cal{N}}(\theta_1,1)$, the likelihood ratio test is $\left\{Z>c/\sqrt{n}\right\}$.

With Pocock's example, under $D \ge 1$, $Z_1$ is a normal random variable used for stage 1 hypothesis testing, and 
$$\log LR(Z_1|D\ge 1) = -n_1\left(\log \theta_0 - \log \theta_1\right) - \sqrt{n_1}Z_1/\theta_1.$$
 Then, 
$$\log LR(Z_{(2)}|D\ge 2) = -(n_1+n_2)\left(\log \theta_0 - \log \theta_1\right) - \sqrt{n_1+n_2}Z_{(2)}/\theta_1 + C,$$
where $C = \log \text{Pr}\left(Z_1<c_1|\theta_0\right) - 
\log \text{Pr}\left(Z_1<c_1|\theta_1\right)$ is independent of data  given $D \ge 2$. By Theorem \ref{theorem1}, $\{Z_{(d)} \ge c_d\}$ is the UMP test at stage $d$ given $\alpha_d$ and $n_d$, and  no other sequential test is more powerful overall.

\section{Impact and Summary} \label{Conclusion}


To establish that a sequence of likelihood ratio tests are most powerful, we began by constructing the joint probability distribution over the set of  \textit{possible} events. The use of an early stopping criterion eliminates the possibility of some realizations of cumulative test statistics $\mathbf{Z}=\left(Z_{(1)},\ldots,Z_{(K)}\right)$.  This makes an otherwise normal random process $\mathbf{Z}$ unobservable. On its true \textit{stopping rule adapted support}, the distribution of $\mathbf{Z}$ is a mixture of truncated distributions at each stage. Thus, unadapted distributional assumptions should be modified to take into account the planned adaptation scheme. \citet{liu1999} and \citet{liu2006} recognized change in support in the one-parameter exponential family  and investigated bias estimation. \citet{schou2013meta} recognized presence of truncation in the joint distribution of stage-specific test statistics, but they mostly focused on bias in meta-analytic studies. The adapted support is critical for derivation MLEs' and test statistics'  distributions in Section \ref{sec:densities}.

In Section \ref{sec:densities}, distinct probability measures are derived for design [unconditional], interim hypothesis testing [conditional on collecting data up to the time of interim testing]
and when the study is completed [conditional on deciding to stop at a particular stage]. These probability distributions formalized a new probabilistic framework (Section \ref{sec:densities}) which is used in Section \ref{mostpower} to show no testing sequence is more powerful than sequential likelihood ratio tests.

Likelihood ratio tests are most powerful for testing simple hypotheses and, under monotone likelihood ratio, for testing composite hypotheses; see \citet{neyman1933} and Karlin-Rubin theorem in \citet{ferguson2014mathematical}. The Karlin-Rubin theorem works with the one-parameter exponential family and Section \ref{mostpower} shows that most powerful testing continues to hold for sequential tests. Even though distributions of sample-size-dependent sufficient statistics (\citet{blackwell1947}) belong to a curved exponential family (\citet{efron1975, liu1999,liu2006}), the  distributions conditional on the sample size still belong to one-parameter exponential family. This fact is used in Section \ref{mostpower} to show that likelihood ratio sequential tests are most powerful tests with any pre-determined $\alpha$-spending function. This result is applicable to many common group sequential designs.

Treatment-effect-dependent (non-ancillary) stopping rules are part of  many common GSDs including Pocock \cite{Pocock1977}, O'Brien \& Fleming \cite{Brien1979}, and Haybittle-Peto designs \cite{haybittle1971,peto1976}. If the data follow a normal distribution with known variance (without possibility of early stopping), then these designs are most powerful for their $\alpha$-spending functions. Similarly, application of Simes test (see \citet{simes1986improved}) and its recently proposed modification for sequential testing (see \citet{Tamhane2020}) cannot have higher power than group sequential tests from  \citet{jennison1999} with the same $\alpha$-spending function.

Historically, many researchers have relied, and currently rely, on joint normality (e.g., \citet{jennison1999}, \citet{Proschan2006}, \citet{Kunz2020}).  Critical  adjustments ate  often done with recursive sub-density estimation; see \citet{Armitage1969}. Section \ref{connectArmitage} places Armitage's sub-density formula in the broader probability framework of possible events and confirms, from the prospective of this $\sigma$-algebra, that the common current practice of using Armitage's formula for calculating distributions of interim test statistics is appropriate. 

There is plenty of evidence that adaptive designs make statistics non-normally distributed. \citet{Demets1994} point out that the distribution of stage-specific test statistics is not normal and should be estimated recursively. \citet{jennison1999} plot the density of a  normal test statistic used in  GSD settings [pages 174-177], where discontinuity points clearly show non-normality. \citet{Li2002} find the joint density of stage 1 and stage 2 standardized test statistics not to be bivariate normal. Local asymptotic non-normality was established following sample size recalculations (SSRs) that depend on an interim observed treatment effect (\citet{Tarima2019,Tarima2020}); and a GSD with a single interim analysis can be viewed as a special case of an SSR. MLEs converging to random mixtures of normal variables have been found in other adaptive designs (\citet{Ivanova2000}, \citet{Ivanova2001}, \citet{May2010}, \citet{Lane2012},
\citet{flournoy2018effects}).

\citet{milanzi2015} developed a likelihood approach that applies when the early stopping rule does not depend on the parameter of interest. In this case, sample size adaptation is ancillary to the treatment effect and  asymptotic normality of MLEs holds.  \citet{gnedenko1996, Bening2012,  christoph2020second, korolev2019asymptotic} assume that the distribution of the random sample size  does not depend on previosly collected data; \citet{molengerghs2012} assumes this asymptotically.

Nevertheless, convergence of sample means to non-normal random variables was shown even for ancillary random sample sizes. In \cite{gnedenko1996}, Gnedenko and Korolev show convergence of standardized sums with random number of summands of infinitely divisible random variables to mixtures of stationary distributions. They give conditions for convergence to a mixture of normal distributions. Bening et al. \cite{Bening2012} and Christoph  et al. \cite{christoph2020second}  explore  convergence to  mixtures of normal distributions and to Student's limit distribution.
When convergence  is \emph{mixed} (see, for example \citet{hausler2015stable}), \citet{lin2020random} shows how norming with the observed information  can result in a normal limit. However, the requirement for mixed convergence appears strong, and it does not cover the limiting mixtures obtained in this paper.

The impact of early stopping is  pervasive.  It affects the probabilistic characterisation of the tests (e.g., type I error and Fisher Information) as well as  the distributions of MLEs and test statistics. Its effect on Fisher information, when stopped at different stages,  is not widely recognized.

During the design phase, before observations are taken,  the full form of the joint density (\ref{densjoint}), accounting for all possible events, is appropriate. This contrasts with the current practice of using the density assuming the experiment will continue through stage $K$.  More details on the differences resulting from these two design approaches will be the subject of another paper.

 

However, normality and asymptotic normality assumptions continue to be directly used with non-normally distributed statistics. We identify two main reasons for this. 
\begin{enumerate}
    \item 
 Many researchers 
consider large sample properties against a fixed treatment effect independent of the sample size. 
From one point of view, a treatment effect is a population quantity which does not change with sample size. But if one develops an asymptotic approximation to the testing environment using a fixed treatment effect, the statistical experiment stops at the first interim analysis with probability one for any consistent test; test statistics degenerate to a point mass; see Section 7.4 in \citet{Fleming1991}. Under a fixed treatment effect, the power converges to one and cannot be used to compare different testing procedures. This issue triggered development of various descriptions of asymptotic relative efficiency. The most popular approach is Pitman asymptotic relative efficiency \cite{pitman1948}, where asymptotic power is evaluated under local alternatives; see \citet{nikitin1995asymptotic}. In addition, local alternatives clearly reflect actual practice for experiment planning. Experiments are never planned for a statistical power = 1. Small sample size studies (pre-clinical, animal studies) are planned to detect large effect sizes, moderate sample sizes (typical phase 3 studies) are used to detect moderate effect sizes, and large sample sizes (epidemiological studies, like vaccine studies) are used to detect small differences.

 \citet{koopmeiners2012} explored MLEs conditional on stopping, but assumed asymptotic normality to evaluate their uncertainty. \citet{martens2018} relied on asymptotic normality for evaluating regression coefficients under the Fine--Gray model in GSD settings. \citet{Asendorf2018} evaluated asymptotic properties with SSR under a fixed alternative  for negative binomial random variables.
\item Some researchers  investigate \textit{local} asymptotic properties when early stopping is not possible or is ancillary to the treatment effect. This also leads to asymptotic normality. \citet{Scharfstein1997} show that without possibility of early stopping ``time-sequential joint distributions of many statistics $\ldots$ are multivariate normal with an independent increments covariance structure'' under local alternatives. These results are generally consistent with the classical results on local asymptotic normality of \citet{lecam1960}, where both mean and variance of the limiting normal distribution depend on the parameter of interest; see Chapter 7 of \citet{vandervaart1998}. However, Section \ref{LSP} shows that the possibility of early stopping destroys local asymptotic normality: the limiting distribution of standardized test statistics is a mixture of truncated normal distributions. Similar findings were previously proved for non-ancillary sample size recalculations; see \citet{Tarima2019}.
\end{enumerate}
\citet{gao2013} is a rare exception  in not making a normality assumption; these authors mostly deal with set operations and probabilities and, using stage-wise ordering of events, they calculate P-values, confidence intervals, and a median unbiased estimate of the parameter of interest.



It is recommended that the full form of the joint density (\ref{densjoint}), accounting for all possible events, be used for study design before observations are taken. This contrasts with the current practice of using the density assuming the experiment will continue through stage $K$.


\section*{Conflict of interest}
 The authors declare that they have no conflict of interest.

\bibliographystyle{Chicago}
\bibliography{bib}

\end{document}